\documentclass[a4paper,fleqn,usenatbib]{mnras}

\usepackage[T1]{fontenc}
\usepackage{ae,aecompl}

\usepackage{graphicx}	
\usepackage{amsmath}	
\usepackage{amssymb}	

\title[Characterization of CO$_2$ ice]{Characterization of thin film CO$_2$ ice through the infrared $\nu_1+\nu_3$ combination mode}

\author[J. He \& G. Vidali]{
Jiao He,\thanks{E-mail: jhe08@syr.edu}
Gianfranco Vidali,\thanks{E-mail: gvidali@syr.edu}
\\
Physics Department, Syracuse University, Syracuse, NY 13244, USA\\
}

\date{Accepted XXX. Received YYY; in original form ZZZ}

\pubyear{2017}

\begin{document}
\label{firstpage}
\pagerange{\pageref{firstpage}--\pageref{lastpage}}
\maketitle
\begin{abstract}
Carbon dioxide is abundant in ice mantles of dust grains; some is found in
the pure crystalline form as inferred from the double peak splitting of the
bending profile at about 650~cm$^{-1}$. To study how CO$_2$ segregates into the
pure form from water-rich mixtures of ice mantles and how it then crystallizes,
we used Reflection Absorption InfraRed Spectroscopy (RAIRS) to study the
structural change of pure CO$_2$ ice as a function of both ice thickness and
temperature. We found that the $\nu_1+\nu_3$ combination mode absorption profile
at 3708~cm$^{-1}$ provides an excellent probe to quantify the degree of
crystallinity in CO$_2$ ice. We also found that between 20 and 30~K, there is
an ordering transition that we attribute to reorientation of CO$_2$ molecules,
while the diffusion of CO$_2$ becomes significant at much higher temperatures.
In the formation of pure crystalline CO$_2$ in ISM ices, the rate limiting
process is the diffusion/segregation of CO$_2$ molecules in the ice instead of
the phase transition from amorphous to crystalline after clusters/islands of
CO$_2$ are formed.
\end{abstract}

\section{Introduction}
CO$_2$, along with H$_2$O and CO, is the main component of the ice mantle
covering interstellar dust grains in molecular clouds. In space, solid CO$_2$
is observed in the mid-infrared by the bending mode ($\nu_2$) at
$\sim$650~cm$^{-1}$ \citep{Gerakines1999, Pontoppidan2008, Ioppolo2013,
Noble2013}, the asymmetrical stretching mode ($\nu_3$) at $\sim$2350 cm$^{-1}$
\citep{Gerakines1999, Nummelin2001,Noble2013}, as well as the combination modes
$\nu_1+\nu_3$ at 3708 cm$^{-1}$ and $2\nu_2+\nu_3$ at 3600 cm$^{-1}$
\citep{Gerakines1999, Keane2001}. $^{13}$CO$_2$ can also be observed via the
$\nu_3$ mode at $\sim$2280 cm$^{-1}$ \citep{deGraauw1996, Boogert2000}.  The
symmetric stretch $\nu_1$ is IR inactive. The abundance of solid CO$_2$ with
respect to H$_2$O ice ranges from 5\% to 40\%, with a median in the 20--30\%
range, in star formation regions \citep{ Boogert2015, Yamagishi2015}.  Because
the $\nu_3$ mode is intense and is often saturated, the $\nu_2$ mode is often
used instead to study the abundance and the physical and chemical environment
of solid state CO$_2$. \citet{Pontoppidan2008} did a comprehensive survey of
the $\nu_2$ mode of CO$_2$ in various young stellar objects (YSOs) and found
that the $\nu_2$ absorption profile can be separated into a few components
representing different chemical and physical environments for CO$_2$. Most
interestingly, the double peak splitting of the bending mode---the so-called
Davydov splitting \citep{Davydov1962}---for pure crystalline CO$_2$ is found in
many lines of sight \citep{Pontoppidan2008, Ioppolo2013, Noble2013}. To use the
double peak splitting as a thermal history probe, it is crucial to study in the
laboratory how CO$_2$ segregates into patches of pure CO$_2$ and then
crystallizes. The segregation of CO$_2$ from CO$_2$:H$_2$O:CH$_3$OH or
CO$_2$:H$_2$O mixtures has been studied in several prior works
\citep{Sandford1990, Ehrenfreund1998, Ehrenfreund1999, Gerakines1999,
Palumbo2000, Hodyss2008, Ioppolo2013, Isokoski2014, Cooke2016}
using transmission FTIR spectroscopy. In these studies, and in the IR studies
of pure CO$_2$ ice (see \citet{Kataeva2015} and references cited therein) the
ice is much thicker than that of actual ice mantles. It is known that
crystallization is characterized by long range order in the solid; in thin
films, it depends on the thickness, with the thinner films often being
amorphous or partially amorphous with nanocrystals \citep{Loerting2009}. It is
more realistic to study the segregation and crystallization in the thickness
range comparable with the thickness of ice mantles (less than a few tens of
a monolayer (ML)) in the ISM\@. However, thin films transmission spectroscopy
is not sensitive enough to give data with high enough signal to noise ratio.
Reflection Absorption InfraRed Spectroscopy (RAIRS) provides a more sensitive
alternative to study the segregation and crystallization of CO$_2$ in thin film
ice mixtures, although in general RAIRS spectra has different absorption
profile than transmission spectra \citep{Baratta1998}. \citet{Oberg2009} used
RAIRS and adopted a more realistic thickness (< 40 ML) of CO$_2$:H$_2$O mixture
to quantify the segregation of CO$_2$. They found that segregation of CO$_2$
becomes significant between 50 and 60~K. Most recently, \citet{He2017} measured
the binding energy of CO$_2$ on water ice and on CO$_2$ ice, and found that the
former is weaker than the latter. Therefore with enough thermal energy CO$_2$
is more likely to bind to other CO$_2$ molecules to form clusters instead of
binding to water. \citet{He2017} obtained the diffusion energy barrier of
CO$_2$ on the surface of compact amorphous solid water. On
a laboratory time scale the diffusion becomes significant at above 60~K, which
is in agreement with the temperature found in prior studies of the segregation
of CO$_2$ from CO$_2$:H$_2$O mixtures \citep[eg.,][]{Hodyss2008, Oberg2009}.

While the segregation of CO$_2$ in water-based mixtures has been studied by
several groups, fewer details are available of the crystallization process of
pure CO$_2$ ice at low temperature. Furthermore, there is some disagreement on
the results and their interpretation. \citet{Escribano2013} used theoretical
modeling and laboratory measurements using both RAIRS and transmission
spectroscopy to study the crystallization of pure CO$_2$ ice. They found that
even for CO$_2$ ice deposited when the substrate is at 8~K, as long as the ice
film is not too thin, the ice is partly crystalline. They also found that when
an amorphous CO$_2$ ice is heated up to 25~K, it becomes crystalline. From 25~K
to the desorption temperature, the ice structure does not change, as inferred
from the fact that there is no change in the infrared spectra in this
temperature range. This is in contradiction with \citet{Gerakines2015} and
\citet{Isokoski2013}. \citet{Isokoski2013} measured high resolution
transmission FTIR spectra of pure CO$_2$ ice deposited at 15~K as well as
during heating up, and found that the $\nu_2$ mode double peak and the
combination modes $\nu_1+\nu_3$ and $2\nu_2+\nu_3$ become narrower and sharper
during heating up from 15~K to 75~K. This indicates that the ordering in the
ice changes during heating up. \citet{Gerakines2015} found that CO$_2$ ice
deposited at 10~K at the rate of 0.1 $\mu$m/hr, or about 200 times slower than
in \citet{Isokoski2013}, does not show any splitting in the bending mode. They
claimed that the missing of splitting is attributed to the fact that their
CO$_2$ ice is amorphous.  This lack of splitting in \citet{Gerakines2015}
disagrees with both \cite{Escribano2013} and \citet{Isokoski2013}. Since these
three groups used different thickness of CO$_2$ ice, it remains a question
whether the difference among them is due to the difference in ice thickness
and/or deposition rate (\citet{Escribano2013} used a slightly slower deposition
rate than \citet{Gerakines2015}). In this work we study systematically how the
CO$_2$ ice crystallization depends on both ice thickness and temperature. We
also look for all signatures of CO$_2$ ice crystallization other than the
splitting of bending mode, hoping to find alternative probes of CO$_2$
crystallinity.

\section{Experimental setup}
The experiments were carried out in a ultrahigh vacuum (UHV) setup located at
Syracuse University. A base pressure of 2$\times 10^{-10}$ torr can be obtained
routinely. At the center of the chamber there is a gold coated copper disk (the
sample) attached to the cold tip of a liquid helium cryostat. A Lakeshore 336
temperature controller with a calibrated silicone diode and a resistance heater
were used to measure and control the temperature in the range of 8 -- 500 K with
an accuracy better than 50~mK. CO$_2$ gas was deposited onto the sample disk
from the background via a UHV precision leak valve. A stepper motor controlled
by a LabVIEW program was used to drive the leak valve. The deposition dose was
calculated by the integration of pressure over time, assuming 1 Langmuir (1 L,
$1\times10^{-6}$ torr$\cdot$s) of exposure is equivalent to 1 monolayer (ML).
Later we use ML and L interchangeably. It is assumed that at a deposition
temperature of 10~K the sticking of CO$_2$ is unity \citep{He2016a}, and the
pressure at the ionization pressure gauge is the same as that in front of the
sample. The standard gas correction factor for CO$_2$ has been taken into
account when calculating the deposition dose. With the automated leak valve,
the deposition rate and dose can be controlled to {\em relative} uncertainty
less than 3\%. The main uncertainty in thickness comes from the pressure reading
of the ion gauge. A systematic error as much as 30\% is possible in this type of
gauges. For experiments in this work, a CO$_2$ deposition rate of 4 L/minute was
adopted except for the deposition of water and CO$_2$ mixture. The following thicknesses were attempted: 1, 2, 5, 10, 15, 20, 30 L.
After deposition at 10~K, the sample was heated up to 100 K at a rate of 0.1~K/s
to desorb the CO$_2$ ice (Temperature Programmed Desorption). The infrared
spectra of the CO$_2$ ice was monitored by a Nicolet 6700 Fourier Transform
InfraRed (FTIR) spectrometer in the Reflection Absorption InfraRed Spectroscopy
(RAIRS) setup with an incidence angle of 78 degrees. The FTIR collects and
averages 7 spectra from 600 cm$^{-1}$ to 4000 cm$^{-1}$ at a resolution of 1
cm$^{-1}$ every 10 seconds, both during deposition and during Temperature
Programmed Desorption (TPD).

\section{Results and Analysis}
\begin{figure*}
  \includegraphics[width=2\columnwidth]{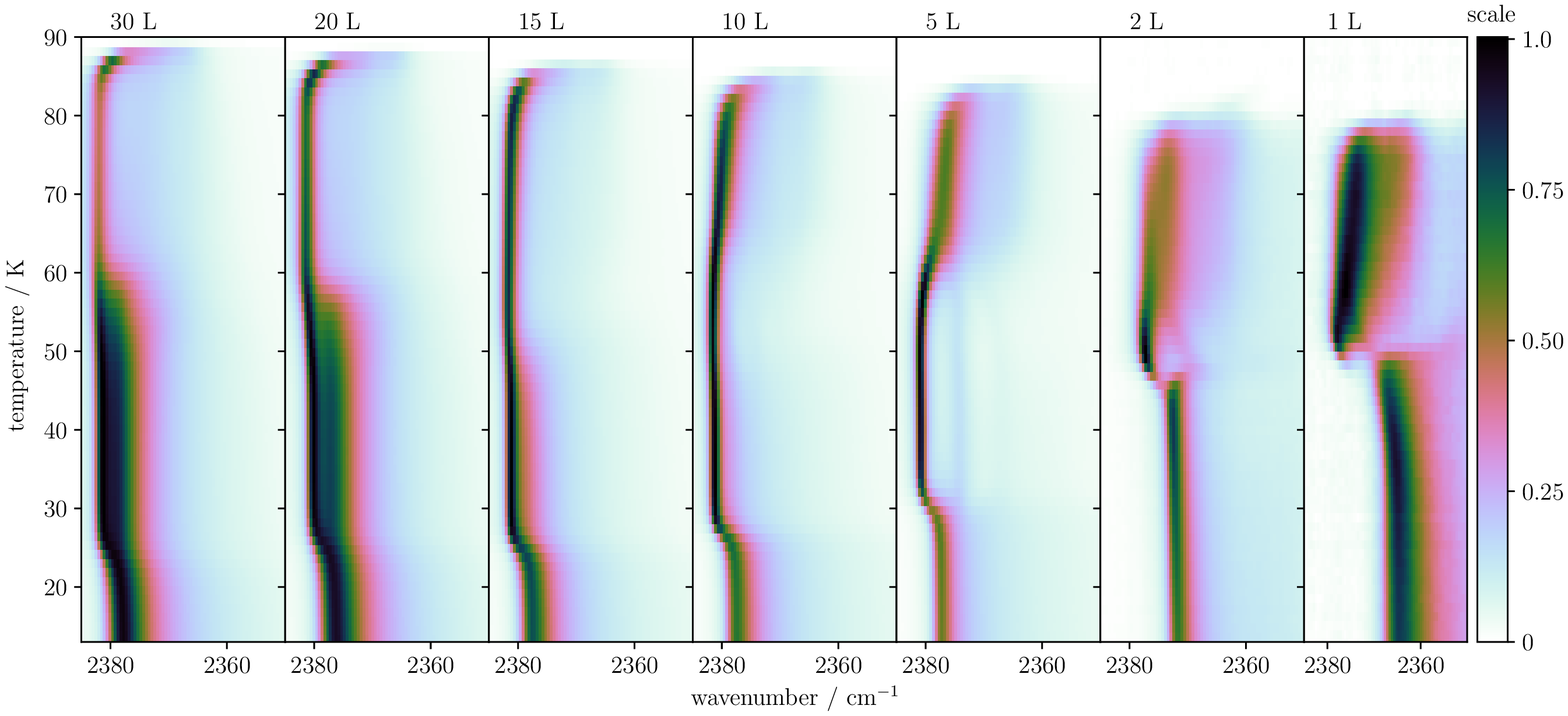}
  \caption{Normalized absorption spectra of the CO$_2$ asymmetrical stretching
    ($\nu_3$) mode during TPDs. The thickness of the CO$_2$ ice is marked on the
    top of each panel. The colorbar scale, on the right side of the figure,
    shows relative intensity.}
\label{fig:color}
\end{figure*}

\begin{figure*}
  \includegraphics[width=2\columnwidth]{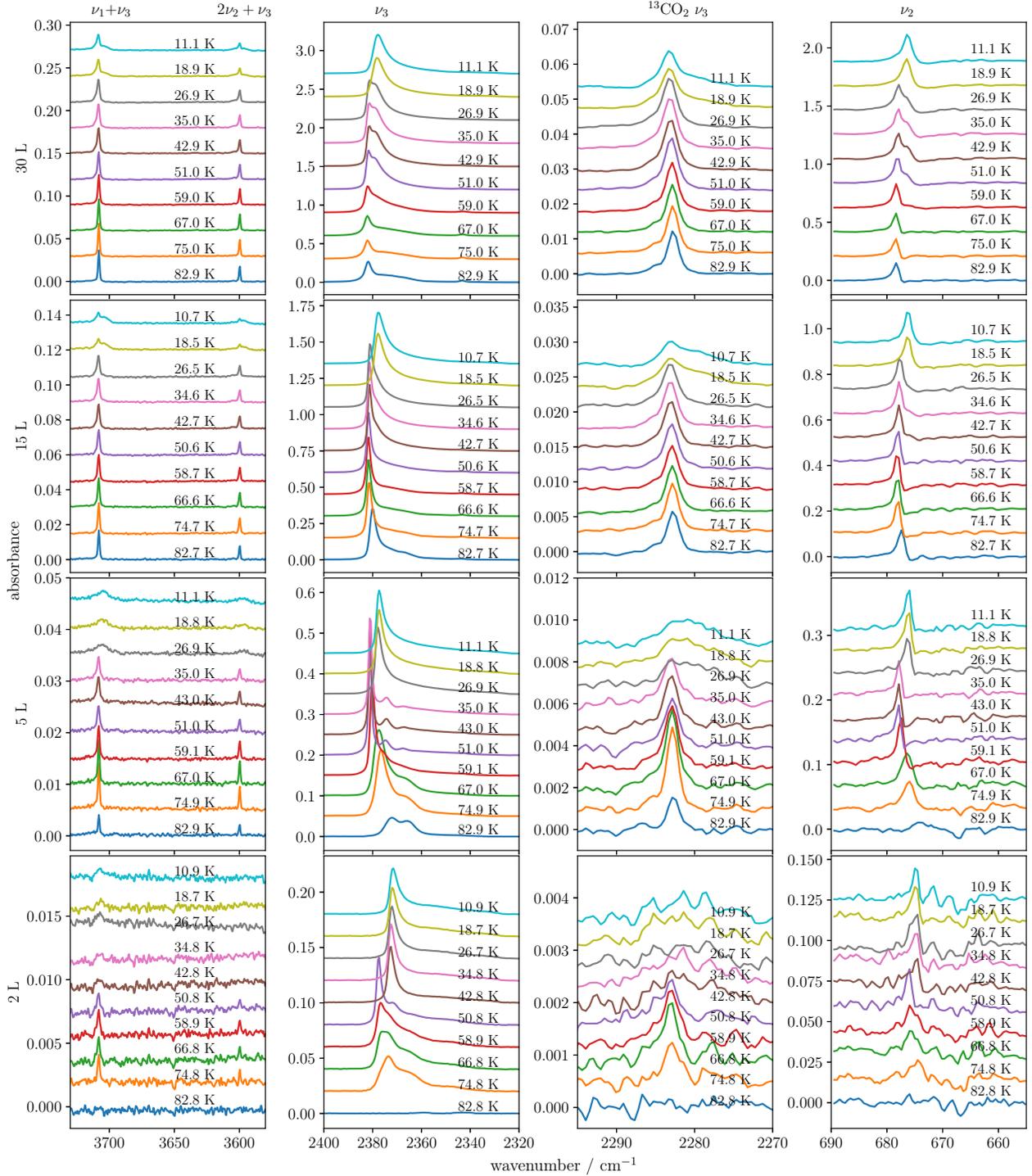}
  \caption{RAIRS of 30, 15, 5, and 2 L of CO$_2$ at selected temperatures. The
    thickness is labeled on the left side of each row, and the vibrational
    modes are labeled on the top of each column. Spectra are offset for
    clarity. The temperature for each curve is also marked. The $\nu_3$ mode
    of $^{13}$CO$_2$ (mixed in $^{12}$CO$_2$ with natural abundance) is also
    shown.}\label{fig:spectra}
\end{figure*}

We deposited 1, 2, 5, 10, 15, 20, and 30 L of CO$_2$ onto the gold surface at
10~K, and then heated up the sample from 10 K to 100 K with a ramp rate of 0.1
K/s. The asymmetrical stretching mode ($\nu_3$) absorption spectra for all these
thicknesses, normalized to the maximum of all the spectra for the same thickness
during TPD, are shown in Fig.~\ref{fig:color}. Fig.~\ref{fig:spectra} shows the
absorption spectra during TPD at selected temperatures for selected thicknesses
(30, 15, 5, and 2 L). The following vibrational modes are shown: $\nu_1+\nu_3$
mode at 3708 cm$^{-1}$, $2\nu_2+\nu_3$ mode at 3600 cm$^{-1}$, asymmetrical
stretching mode $\nu_3$ at $\sim$2380 cm$^{-1}$, $\nu_3$ mode for $^{13}$CO$_2$
at $\sim$2280 cm$^{-1}$, bending mode $\nu_2$ at $\sim$675 cm$^{-1}$. For
$\nu_3$ and $\nu_2$ modes, the position of the absorption peaks are blue shifted
respect to typical spectra measured in transmission. This is due to
the splitting of transverse optical (TO) and longitudinal optical (LO) modes. In
RAIRS measurement, usually the LO mode is seen, while in transmission setup at
normal incidence only TO mode is excited. For polycrystalline films, in the
reflection mode at normal incidence, a small LO peak is seen as well
\citep{Kataeva2015}. The LO and TO modes are present both in amorphous and
crystalline solids \citep{Berreman1963}.

\begin{figure}
  \includegraphics[width=1\columnwidth]{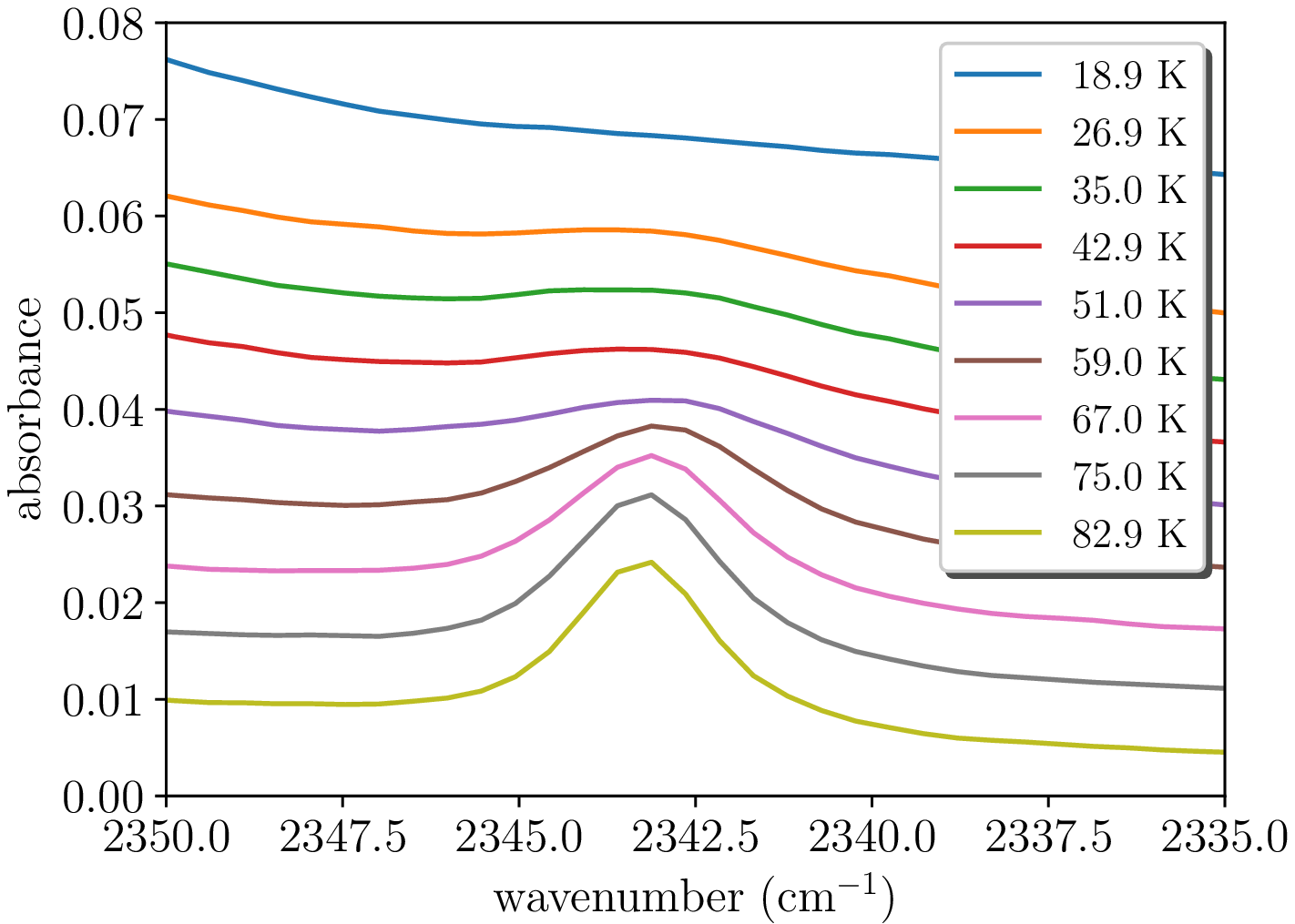}
  \caption{Zoom-in of the region of CO$_2$ $\nu_3$ TO mode for 30~L CO$_2$ ice
  during TPD at selected temperatures.} \label{fig:TO}
\end{figure}

We first analyze the $\nu_3$ mode. In Fig.~\ref{fig:spectra} the second column
shows the $\nu_3$ profile for different temperatures. For all the thicknesses,
at the lowest temperature the peak is centered at below 2380 cm$^{-1}$, typical
of LO mode of disordered CO$_2$. Depending on the thickness, at different
temperatures the peak blue shifts to 2381 cm$^{-1}$, which is the typical
position for crystalline CO$_2$ LO mode \citep[eg.,][]{Escribano2013}. As seen
in Fig.~\ref{fig:color}, for 30~L of CO$_2$, the first change happens at
$25\pm2$~K. In thinner ices, this temperature at which there is a change is
higher. For 1~L of CO$_2$, it is at $\sim$50~K. For the 30~L and 20~L ices, at
above $\sim$60~K the peak becomes much narrower and the red wing almost
disappears. This indicates the formation of long-range order (complete
crystallization), as confirmed by the behavior of the $\nu_1+\nu_3$ combination
mode to be discussed below. The sensitivity to the macroscopic environment of the $\nu_1+\nu_3$ combination
mode was noted before for nanoparticles \citep{Bauerecker2005}.  At the same time, the overall peak area of $\nu_3$
becomes much smaller. This can be explained if the orientation of CO$_2$
molecules in the crystal are in a configuration that lowers the LO absorption.
If we assume in the amorphous phase the average angle between the linear CO$_2$
molecules and the surface normal is 45 degrees, in the crystalline form this
angle is likely to be smaller than 45 degrees, and the LO mode (which excite
vibration mode in the normal direction of the substrate surface) of
asymmetrical stretching becomes less intense. At the same time we should expect
to see an increase in TO mode absorption. Indeed, Fig.~\ref{fig:TO} shows the
increase in TO mode absorption at 2343 cm$^{-1}$ at $\sim$60~K. In the RAIRS,
TO mode absorption is usually much weaker than LO because TO mode cancels out
at the metal surface if the incident angle is close to 90 degrees. Another
dramatic change happens above 80~K during the desorption of CO$_2$ ice. The
peak red shifts from $\sim$2381 cm$^{-1}$ to lower than 2380 cm$^{-1}$, and it
broadens. This may be because the symmetry is broken and the ice becomes disordered
again, due to the rapid movement of CO$_2$ molecules during desorption.

In other vibration modes, there are also significant changes accompanying the
structural changes. In the $\nu_2$ LO mode, amorphous CO$_2$ shows a peak at
$\sim$676~cm$^{-1}$; as the ice crystallizes, it blue shifts to $\sim$679
cm$^{-1}$. The presence of the naturally occurring isotopic impurity $^{13}$CO$_2$ provides another powerful marker of the morphology of the film. The $\nu_3$ mode of $^{13}$CO$_2$ shows a change from a broad band to a narrow peak at the same temperature of the onset of crystallization that was mentioned above about  the  $\nu_3$ mode of $^{12}$CO$_2$. Because of the low temperature  at the transition ($\sim$ 25~K and higher for thinner films) there is no diffusion at the laboratory time scale. Therefore, single $^{12}$CO$_2$ molecules remain in place, but their environment changes, and this is reflected by the sharpening of the $\nu_3$ mode. The peak sharpens with
increasing degree of crystallinity while the position remains the same at 2283
cm$^{-1}$. The combination modes $\nu_1+\nu_3$ and $2\nu_2+\nu_3$ have the
same shape and their magnitudes are proportional. In the ice deposited at low
temperature and especially for the thinner films, the combination modes show a
broad profile, as expected for an amorphous solid. Here we use the word
amorphous loosely, since there is no direct information on the degree of
ordering of the deposit. As an intermediate range order \citep{Price1996} structure
forms, a sharp component shows up, and becomes more pronounced as longer
range order is formed. The degree of crystallinity is strictly positively
related to the magnitude of the sharp component. We suggest that these
combination modes are excellent probe of CO$_2$ ice crystallinity because they
are very pronounced and easy to separate from the broad amorphous component.

Singling out one sharp component is numerically much easier than singling out
a double peak. Another advantage of using the combination modes as
a crystallinity probe is that they show the same behavior in both
transmission and RAIRS spectra \citep{Bauerecker2005}. On the surface of dust
grains the combination modes of CO$_2$ can also be directly compared with
RAIRS/transmission measurements without grain shape corrections.  To study the segregation of CO$_2$ in ice
mantle, thin ice layers are preferred and RAIRS is more sensitive for such
measurement. The $\nu_2$ and $\nu_3$ modes in RAIRS are dramatically different
from the transmission spectra. Combination modes provide a consistent profile
regardless of the geometry of laboratory setup. The coincidence of sharp combination modes and
bending mode splitting in transmission spectra has been observed in prior
studies \citep{Ehrenfreund1998, Hodyss2008}, which verifies the validity of
using combination modes as indication of crystallization, and also supports
that this indication of crystallization applies to transmission spectra as
well.

\begin{figure}
  \includegraphics[width=1\columnwidth]{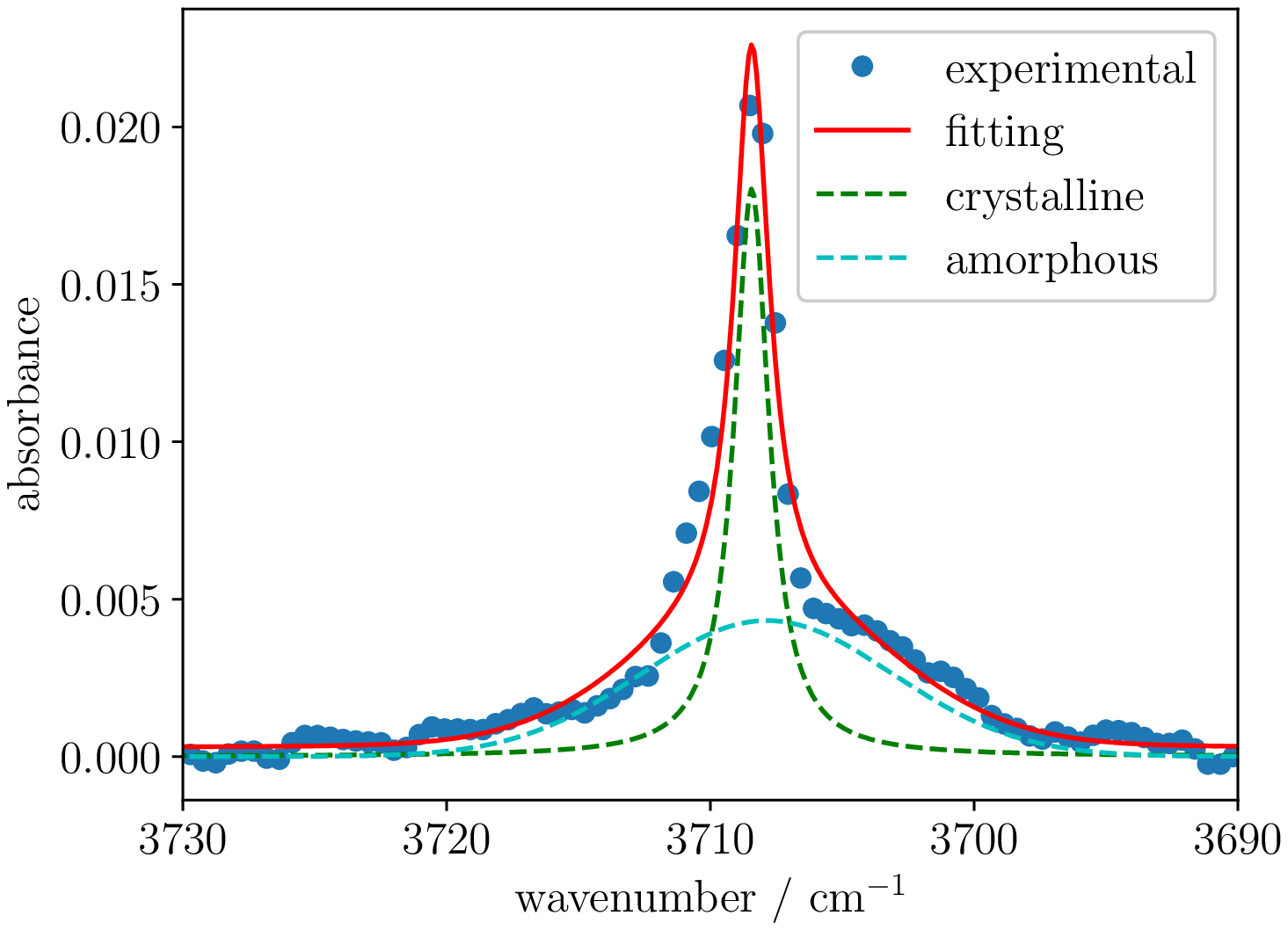}
  \caption{A typical fitting of the $\nu_1+\nu_3$ combination mode profile
  using Lorentzian (cyan) and Gaussian (green) distributions.}
\label{fig:fit}
\end{figure}

\begin{figure}
  \includegraphics[width=1\columnwidth]{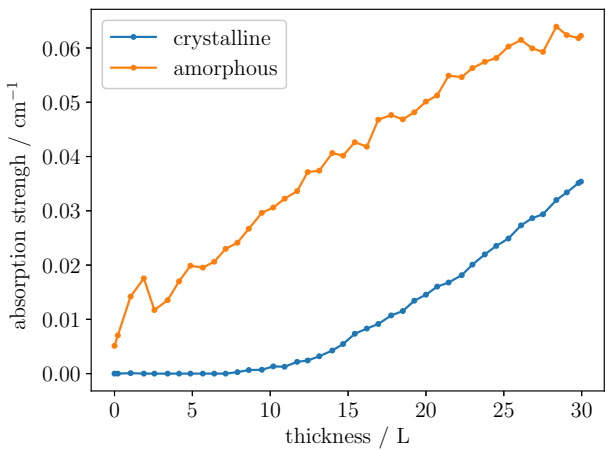}
  \caption{The area of Lorentzian (crystalline) and Gaussian (amorphous)
  components during the deposition of 30~L CO$_2$ on gold substrate at
10~K.}\label{fig:30L}
\end{figure}

Now we show how the $\nu_1+\nu_3$ mode can be used as a probe to study in detail
the crystallization of pure CO$_2$ ice. The $\nu_1+\nu_3$ absorption peak can be
fit using two components, one sharp and narrow component centered at 3708
cm$^{-1}$ best fitted with a Lorentzian distribution with $\gamma=0.7$
cm$^{-1}$, one broad component fitted with a Gaussian distribution with
$\sigma>4$ cm$^{-1}$. The Lorentzian component is the signature of crystalline
CO$_2$ while the Gaussian component represents amorphous (or disordered) CO$_2$
ice. In a single crystalline CO$_2$ ice only the Lorentzian component can be
seen while in the other extreme, fully amorphous CO$_2$ ice has only a broad
Gaussian component. A typical fitting is shown in Fig.~\ref{fig:fit}.
Fig.~\ref{fig:30L} shows the area of both components during the deposition of
the 30~L of CO$_2$ ice when the substrate is at 10~K. Below 10~L the ice is
amorphous. As the thickness increases further the crystalline component
emerges. This suggests that the underlying polycrystalline/amorphous gold
surface may affect the growth of crystalline CO$_2$ ice. An alternative
explanation is that partial crystallization requires at least intermediate
range order, and too thin a film at 10~K can not form the intermediate range
order at 10~K. The result in Fig.~\ref{fig:fit} can be extended to thicker
ices. \citet{Isokoski2013} deposited 3000~ML of CO$_2$ when the substrate was
at 15~K, and the ice is polycrystalline, which agrees qualitatively with
Fig.~\ref{fig:fit}.

\begin{figure}
  \includegraphics[width=1\columnwidth]{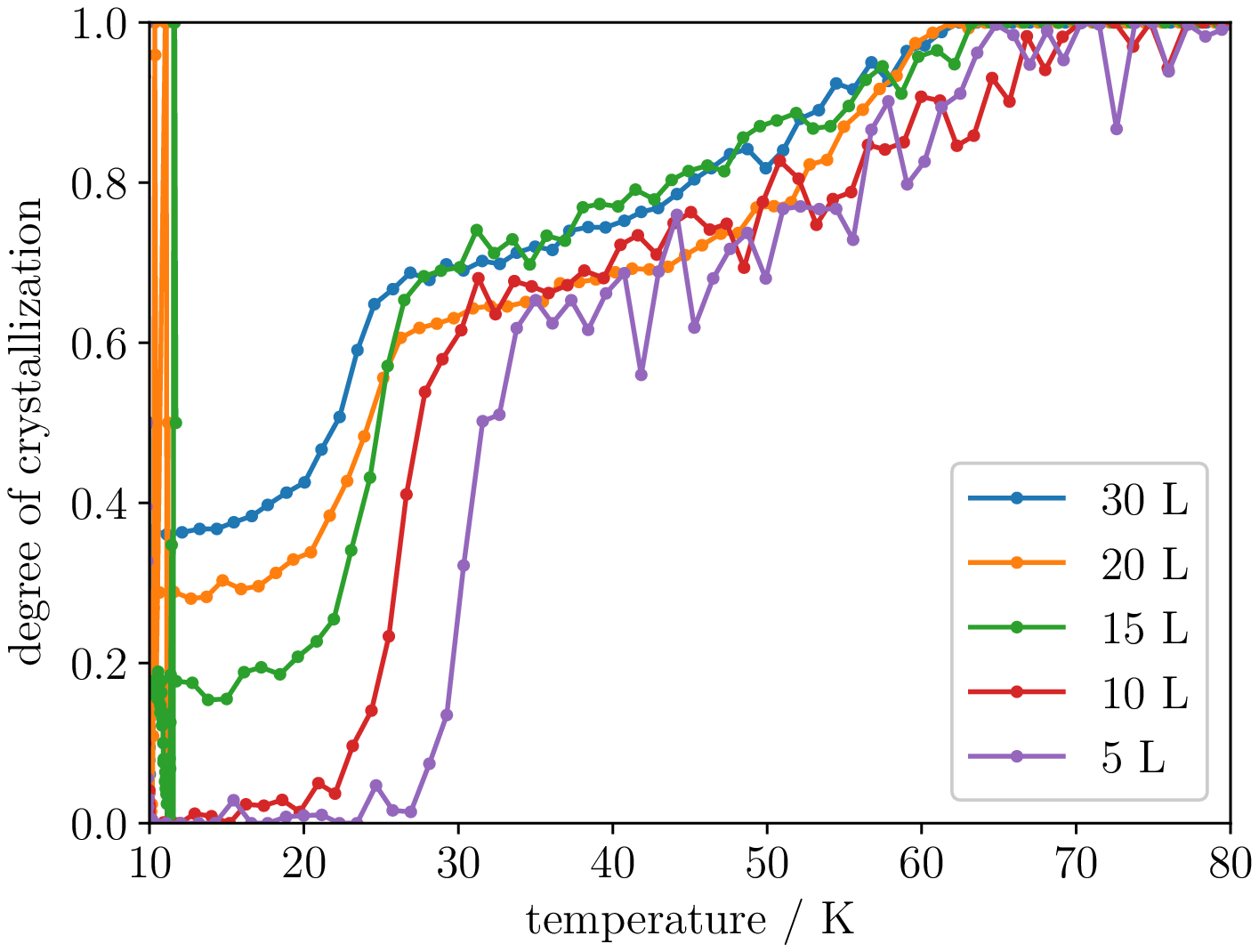}
  \caption{Degree of crystallinity (see text) in CO$_2$
  ices of different thickness during TPD.}\label{fig:crys}
\end{figure}

In order to better quantify the crystallinity, we define a new
parameter---degree of crystallinity ($DOC$) as:
\begin{equation}
DOC = \frac{A_{crystalline}}{A_{crystalline}+A_{amorphous}}
\end{equation}
where $A_{crystalline}$ and $A_{amorphous}$ are the absorption strength of the
Lorentzian component and Gaussian component, respectively.  Fig.~\ref{fig:crys}
shows the $DOC$ as a function of temperature during TPD for different thickness
CO$_2$ ices. In thinner than 5~L ices, the combination modes are too noisy and
we don't quantify the $DOC$. There is a dramatic increase in $DOC$ between 20
and 30~K for all of the thicknesses, which suggests a significant crystallinity change in
this temperature range. We attribute this change to orientational changes of CO$_2$
molecules in this temperature range: CO$_2$ molecules reorient and forms
intermediate range order (nanocrystals). There are few studies on orientational
ordering in CO$_2$ cubic ice \citep{Torchet1996,Kuchta1988,Krainyukova2017}.
Solid, crystalline CO$_2$ ice has Pa$\bar{3}$ symmetry in vacuum with four
molecules per elementary cell placed along the cubic diagonals. The molecules
perform small librations around the diagonal (with the C atom on the diagonal)
\citep{Kuchta1988}. A recent THEED (Transmission High Energy Electron
Diffraction) study on thin ($\sim$10~nm) films deposited at $\sim$65~K reveals
a more complicated rotational motions with the molecular tips (the oxygen
atoms) hopping in 24 equivalent positions with a maximum deviation from the
diagonal of about 30 degrees and decreasing from 15 to 70~K
\citep{Krainyukova2017}.

In the growth of 30 and 20~L ices, although during deposition the sample is at
10~K, gas-phase CO$_2$ molecules possess room temperature thermal energy, and
therefore CO$_2$ molecules can reorientate right after landing on the surface
and form nanocrystals. For thinner ices, it is more difficult to form
structures with long-range order. In the extreme case where the ice is very
thin (1--2~L), molecules have to diffuse on the substrate surface and form
clusters or islands before an intermediate range order (nanocrystalline) structure
can be formed. From Fig.~\ref{fig:color} it can be seen that for 1~L of CO$_2$
the transition temperature is at 50~K, which is close to the temperature of
diffusion of CO$_2$ on a compact water ice surface \citep{He2017}.  In very thin CO$_2$ ices,
the crystallization temperature is mostly limited by diffusion. The same
conclusion should also apply for low concentration of CO$_2$ mixed in water
ice. For ice thicker than 5~L, at higher temperatures, the $DOC$ value
increases gradually to unity at about 65~K, and ice becomes almost all
crystalline. The emerging $\nu_3$ TO mode absorption at about 60~K
(Fig.~\ref{fig:TO}) also supports the formation of long range order.

\section{Discussion}
\begin{figure}
  \includegraphics[width=1\columnwidth]{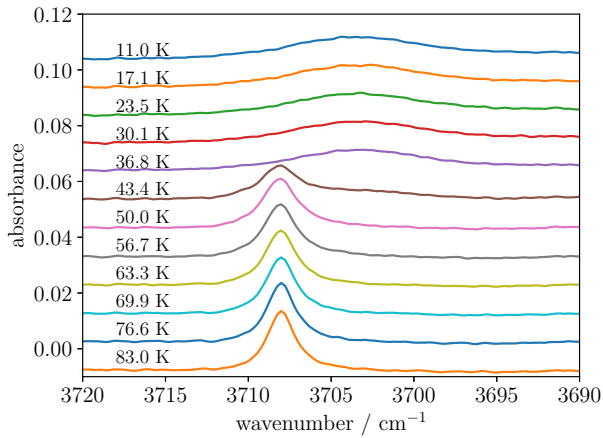}
  \caption{$\nu_1+\nu_3$ mode of 22 ML of 1:10 H$_2$O:CO$_2$ mixture deposited
  at 10 K and heated up at 0.1 K/s. The temperature for each spectrum is labeled.
  }\label{fig:mix}
\end{figure}

\begin{figure}
  \includegraphics[width=1\columnwidth]{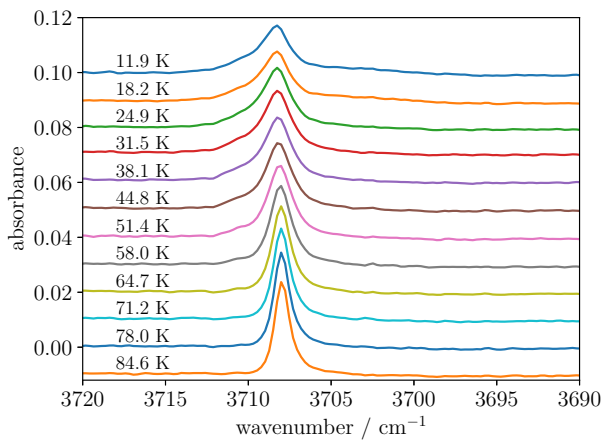}
  \caption{$\nu_1+\nu_3$ mode of 20 ML of pure CO$_2$ ice deposited at 10 K and
  heated up at 0.1 K/s. The temperature for each spectrum is labeled.
}\label{fig:control}
\end{figure}

The infrared spectra of pure CO$_2$ ice deposited at low temperature has been
measured previously, e.g. \citep{Sandford1990, Edridge2013, Escribano2013,
Isokoski2013, Gerakines2015}; here we compare our results with some of the
prior studies. In \citet{Isokoski2013}, Fig. 5 shows that the 3000~ML ice grown
at 15~K already demonstrates a certain degree of crystallinity and is in the
polycrystalline form, in agreement with our results. Between 15~K and 75~K, the
degree of crystallinity increases. At 75~K the ice is likely fully
crystallized. In \citet{Escribano2013}, Fig. 2 shows that the double peak
splitting of the bending mode emerges between 20~K and 25~K; this also agrees
with our $DOC$ vs.\ temperature curve. However, Fig. 2 in \citet{Escribano2013}
shows almost no difference in the bending profile above 25~K until desorption of the
ice. This is different from our measurement and \citet{Isokoski2013}'s.
It is probably due to the relatively low signal to noise ratio in
\citet{Escribano2013}; therefore small changes in the absorption profile can
not be recognized. Fig. 1 of \citet{Escribano2013} also shows an increase in
degree of crystallinity during deposition when the ice is still thin, in
agreement with Fig.~\ref{fig:30L} of our work.  In Fig. 6 of \citet{Escribano2013} the 200 ML of CO$_2$ deposited at 14~K already shows some character of crystalline ice, and this also agrees with our results. In \citet{Edridge2013}'s
experiments, CO$_2$ was deposited at 28~K, which is about the transition
temperature from the amorphous phase. Their $\nu_3$ profiles show a sharp peak
at all temperatures during TPD, indicating (poly)crystalline structure.
\citet{Gerakines2015} also studied the phase of CO$_2$ ice, and found that
their CO$_2$ ice deposited at a rate of 0.1 $\mu$m hr$^{-1}$ (200 times slower
than in \citep{Isokoski2013}) has no sign of
crystallization but doubling this deposition rate yields crystalline CO$_2$.
This disagrees with our work as well as with \citet{Escribano2013}'s and
\citet{Isokoski2013}'s experiments. Given the standard methods used in all
these experiments, one is led to conclude that perhaps the explanation lies in
the non ultra-high vacuum conditions used in \citet{Gerakines2015}'s
experiments. This discrepancy could be due to contamination from background
water in the vacuum chamber during CO$_2$ deposition. The effect of water in
CO$_2$ ice has been shown to wipe out the sharp features of both $\nu_3$ and
$\nu_2$ (e.g., \citet{Cooke2016}). A clue that this might be the case resides in
\citet{Gerakines2015}'s experiments when, by increasing the deposition rate
(and, thus, decreasing the deposition time and degree of contamination), the
amorphous-like features (broad $\nu_3$ profile and unresolved $\nu_2$ splitting)
change into crystalline-like ones (sharp $\nu_3$ and $\nu_2$ split).  To
  check the effect that water has on the CO$_2$ ice spectrum, we performed an
  experiment with a 1:10 mixture of  H$_2$O:CO$_2$ at a CO$_2$ deposition rate
  of 1 ML/minute. The result is shown in  Figure~\ref{fig:mix}. Figure~\ref{fig:control} shows the same $\nu_1+\nu_3$ mode  for pure CO$_2$ ice.
  Clearly the mixture with water doesn't show any sign of being crystalline
  below $\sim$40~K, while the pure CO$_2$ ice shows a sharp peak from 10~K.
  When heating it, the mixture with water crystallizes. This agrees with
  \citet{Gerakines2015}'s finding that annealing to 70 K crystallizes the ice.
  Therefore, we suggest that ultrahigh vacuum conditions are important for the
  study of CO$_2$ crystallization.

\section{Astrophysical Implications}
The segregation and crystallization of CO$_2$ in ice mantles is a useful probe
of the thermal history of ices in dense molecular clouds. While prior works
analyzed the double peak splitting of bending profile in efforts to study the
segregation and crystallization CO$_2$ in CO$_2$:H$_2$O mixtures
\citep{Pontoppidan2008,Ioppolo2013,Noble2013} the splitting profile suffers
from the difficulty in separating the double peak feature from CO$_2$ in other
environments. In addition, the shape of the bending mode profile is affected by
grain shapes. We propose that the $\nu_1+\nu_3$ combination mode profile is
a more sensitive probe of pure crystalline CO$_2$ and, because of its weakness,
is not sensitive to dust grain shape effects as the other transitions are
\citep{Keane2001}. Another advantage of the $\nu_1+\nu_3$ combination mode
profile is that it is similar in both transmission spectra and reflection
spectra, therefore can be used by RAIRS measurements and provides more
sensitive measurements of ice films with a thickness comparable with the ice
mantle.  Although very few previous observations of ice mantles provide enough
high resolution data covering the combination mode at 2.7 $\mu$m
\citep{Keane2001}, JWST will provide high quality spectrum data covering this
wavelength region. More reliable information of the thermal history of ice
mantles can be expected from comparing observational spectra with laboratory
measurements.

We also found that there is a dramatic increase in the degree of
crystallinity between 20 and 30~K, and we attribute it to orientational
ordering of CO$_2$ molecules. We introduce a new parameter---degree of
crystallinity (DOC)---to reliably quantify crystallinity of CO$_2$ ice. Above
30~K, the CO$_2$ ice further crystallizes until it becomes fully
crystallized at about 65~K. We can separate the formation of pure crystalline
CO$_2$ in ice mantles into two processes: 1) segregation/diffusion of CO$_2$ to
form clusters/islands of CO$_2$; 2) the formation of long range order in CO$_2$
clusters/islands. The latter process becomes efficiently at above 20~K, while
the former process has been found to happen at a much higher temperature
\citep{He2017,Oberg2009,Hodyss2008}. We therefore concludes that the rate
limiting process is the segregation of CO$_2$ from the CO$_2$:H$_2$O mixture
instead of the crystallization process itself. In the case of a low fraction of
CO$_2$ in the mixture, the diffusion of CO$_2$ on water surface, which has been
studied in detail by \citet{He2017}, becomes the most important process in the
formation of pure crystalline CO$_2$.

\section*{Acknowledgements}
We thank SM Emtiaz, Yujia Huang, and Francis Toriello for technical assistance.
This research was supported by NSF Astronomy \& Astrophysics Research Grant
\#1615897.




\bibliographystyle{mnras}

\bsp	
\label{lastpage}
\end{document}